\documentclass{ptephy_v1}

\usepackage{graphicx}
\usepackage{bm}

\setDOI{ptz}{087}

\def\e{\mathrm{e}}
\def\n{\mathrm{n}}
\def\p{\mathrm{p}}
\newcommand{\bracket}[1]{\big<{#1}\big>}
\begin{document}

\title{Kompaneets equation for neutrinos: Application to neutrino heating in supernova explosions}

\author[1,2,*]{Yudai Suwa}
\affil{Department of Astrophysics and Atmospheric Sciences, Faculty of Science, Kyoto Sangyo University, Motoyama, Kamigamo, Kita-ku, Kyoto 603-8555, Japan}
\affil[2]{Center for Gravitational Physics, Yukawa Institute for Theoretical Physics, Kyoto University, Kitashirakawa Oiwakecho, Sakyo-ku, Kyoto 606-8502, Japan 
 \email{suwa@yukawa.kyoto-u.ac.jp}}

\author[3,4]{Hiroaki W. H. Tahara}
\affil[3]{Research Center for the Early Universe (RESCEU), Graduate School of Science, The University of Tokyo, Tokyo 113-0033, Japan}
\affil[4]{Department of Physics, Graduate School of Science, The University of Tokyo, Tokyo 113-0033, Japan}

\author[5,6]{Eiichiro Komatsu}
\affil[5]{Max-Planck-Institut f\"ur Astrophysik,   Karl-Schwarzshild-Str. 1, D-85741 Garching, Germany}
\affil[6]{Kavli Institute for the Physics and Mathematics of the Universe (Kavli IPMU, WPI), Todai Institutes for Advanced Study, the University of Tokyo, Kashiwa 277-8583, Japan}

\date{Received April 10, 2019; Revised June 15, 2019; Accepted July 3, 2019; Published August 23, 2019}

\begin{abstract}
We derive a ``Kompaneets equation'' for neutrinos, which describes how
the distribution function of neutrinos interacting with matter
deviates from a Fermi-Dirac distribution with zero chemical
potential. To this end, we expand the collision integral in the
Boltzmann equation of neutrinos up to the second order in energy transfer between matter and neutrinos. The distortion
of the neutrino distribution function changes the rate at which
neutrinos heat matter, as the rate is proportional to the mean square
energy of neutrinos, $E_\nu^2$. For electron-type
neutrinos the enhancement in $E_\nu^2$ over its
thermal value is given approximately by
$E_\nu^2/E_{\nu,\rm thermal}^2=1+0.086(V/0.1)^2$ where $V$ is the bulk velocity of
nucleons, while for the other neutrino species the enhancement is
$(1+\delta_v)^3$, where $\delta_v=mV^2/3k_BT$ is the kinetic energy
of nucleons divided by the thermal energy. This enhancement has a
significant implication for supernova explosions, as it would aid
neutrino-driven explosions.
\end{abstract}

\maketitle

\section{Introduction}
\label{sec:intro}

The Boltzmann equation is ubiquitous in physics. In a system in which
matter and radiation interact, a useful approximation can be obtained
by expanding the collision integral in the Boltzmann equation up to
second order in energy transfer between matter and radiation. 

One example is the Kompaneets's equation \citep{komp57}, which
describes how the distribution function of photons evolves via interaction with thermal electrons in the non-relativistic limit. A solution to the Kompaneets equation in the
optically-thin limit is known as the thermal Sunyaev-Zel'dovich (tSZ)
effect \citep{suny72}, which describes a distortion of the black-body
spectrum of the cosmic microwave background photons by inverse-Compton
scattering off hot electrons in galaxy clusters. The tSZ effect has
been routinely detected towards ${\mathcal O}(10^3)$ galaxy clusters
\citep{plan15}.  Another example is the Fermi acceleration
\citep{ferm49}. The so-called second-order Fermi acceleration distorts
the distribution of charged particles by stochastic acceleration due
to time-dependent electromagnetic fields. Both examples can be
formulated in the same form, namely a diffusion equation for the
distribution function of photons or charged particles in momentum
space.

In this paper, we apply the same approximation to the Boltzmann
equation describing neutrinos interacting with matter.  Specifically,
we consider a system in which isotropic neutrinos interact with
nucleons, and expand the collision integral in the Boltzmann equation
up to second order in energy transfer between neutrinos and nucleons. We do not assume that the system is
optically thin. By solving this Kompaneets-like equation for neutrinos,
we obtain distortions of the neutrino distribution function from a
Fermi-Dirac distribution with zero chemical potential as a function of
the temperature and bulk velocity of nucleons.  Our result has a
significant implication for neutrino-driven supernova explosions, as
the distortion of the distribution function changes the rate at which
neutrinos heat nucleons.

The rest of this paper is organized as follows. In
Section~\ref{sec:kompaneets}, we derive the Kompaneets-like equation for
neutrinos interacting with nucleons; the nucleon motion includes both thermal and bulk motion. We present solutions of this equation in Section
\ref{sec:solution}, including the effect of opacity of electron-type neutrinos.
We summarize our results and discuss
their implications in Section \ref{sec:summary}.
In Appendix~\ref{sec:matrix} we review the matrix element of neutrino-nucleon scattering. In Appendix~\ref{sec:rybicki} we provide an alternative derivation of the main result of this paper following the argument of Ref. \citep{rybi79}. Throughout the paper we shall set the speed of light to be unity, $c=1$.

\section{Kompaneets equation for neutrinos}
\label{sec:kompaneets}

We follow derivations of the Kompaneets equation for photons interacting with electrons given in Refs.~\citep{hu94,dode95}, and derive a similar equation for neutrinos interacting with nucleons. To this end, we expand the collision integral of the Boltzmann equation up to second order in energy transfer. We shall ignore the mass of neutrinos throughout this
paper, as the typical neutrino temperature that we consider here (e.g. that in the supernova engine) is much greater than the
current upper bound on the mass of neutrinos on the order of 1~eV. 

The Boltzmann equation is 
\begin{align}
\frac{df(\bm{p},t)}{dt}=&
{\sum_{N}}
\frac1{2p}
\int\frac{d^3q}{(2\pi)^32E_N(\bm{q})}\int\frac{d^3q'}{(2\pi)^32E_N(\bm{q}')}
\int \frac{d^3p'
}{(2\pi)^32p'}
\delta^{(4)}(p+q-p'-q')|M_N|^2
\nonumber\\
&\times\{
g_N({\bm q}')f(\bm{p}',t)[1-f(\bm{p},t)]
-g_N({\bm q})f(\bm{p},t)[1-f(\bm{p}',t)]
\},
\end{align}
where $N$ denotes nucleons (neutrons and protons), $d/dt$ is the Lagrangian time derivative along the
trajectory of a phase-space volume element, $|M_N|^2$ is the spin-averaged matrix element of neutrino-nucleon scattering, 
and $f(\bm{p},t)$ and $g_N({\bm q})$ are the  distribution functions of neutrinos and nucleons with
three-momenta $\bm{p}$ and ${\bm q}$, respectively. 
The four-dimensional Dirac delta function $\delta^{(4)}(p+q-p'-q')$ ensures energy and momentum conservation.

We assume that nucleons are non-relativistic, i.e.~$E_N(\bm{q})\equiv \sqrt{m_N^2+|\bm{q}|^2}\approx m_N+|\bm{q}|^2/(2m_N)$, with the distribution function given by
\begin{equation}
g_N({\bm q})=n_N(2\pi m_N k_BT_N)^{-3/2}\exp\left[-\frac{({\bm q}-m_N {\bm v}_N)^2}{2m_Nk_BT_N}\right]\,,
\end{equation}
where $n_N$, $m_N$, $T_N$ and $\bm{v}_N$ are the number density, mass, temperature and bulk velocity of nucleons, respectively, and $k_B$ is the Boltzmann constant.

Performing  integration over $\bm{q}'$, we obtain
\begin{align}
\frac{df}{dt}=&
{\sum_{N}}
\frac1{8\pi p}\int dp'p'\frac{d\Omega'}{4\pi}
\int\frac{d^3q}{(2\pi)^3}
\frac{|M_N|^2}{E_N(\bm{q})E_N(\bm{q}+\bm{p}-\bm{p}')}\delta[p+E_N(\bm{q})-p'-E(\bm{q}+\bm{p}-\bm{p}')]
\nonumber\\
&\times\{
g_N({\bm q}+\bm{p}-\bm{p}')f(\bm{p}',t)[1-f(\bm{p},t)]
-g_N({\bm q})f(\bm{p},t)[1-f(\bm{p}',t)]
\}.
\label{eq:boltzmann}
\end{align}
The energy transfer is on the order of $E_N(\bm{q})-E_N(\bm{q}+\bm{p}-\bm{p}')\approx (\bm{p}'-\bm{p})\cdot\bm{q}/m_N$, which should be much smaller than $E_N(\bm{q})\approx |\bm{q}|^2/(2m_N)$. 

The next step is to expand $|M_N|^2$, $E(\bm{q}+\bm{p}-\bm{p}')$, the delta function and the distribution functions up to the second order in  energy transfer. In this paper we consider elastic neutrino-nucleon scattering, i.e. $\nu+N\to \nu+N$, and ignore other processes.    

There are three types of the so-called ``neutrinosphere,'' which are determined by different micro processes \citep{raff01}:
\begin{itemize}
\item {\bf Number sphere}: The optical depth of emission and absorption of neutrinos is of order unity. For electron-type neutrinos $\nu_\e$ and
  $\bar\nu_\e$, the processes of electron/positron capture and its inverse processes
  (i.e. $\nu_\e + \n \leftrightarrow \p+\e^-$ and $\bar\nu_\e + \p
  \leftrightarrow \n+\e^+$) are important. For other two types of neutrinos $\nu_\mu, \nu_\tau,\bar\nu_\mu,\bar\nu_\tau$, the pair production/annihilation processes (i.e.
  $\nu\bar\nu\leftrightarrow\gamma\gamma$, $\nu\bar\nu\leftrightarrow
  \e^+\e^-$, $NN\leftrightarrow NN\nu\bar\nu$) determine the opacity.
\item {\bf Energy sphere}: Inelastic scattering by electrons is important here. Electrons receive energy from neutrinos, as the electron rest mass energy (511 keV) is much smaller than the typical neutrino energy ($\sim 10$ MeV), which is determined by the matter temperature at the number sphere. Inside the energy sphere the neutrinos are thermalized due to energy transfer with electrons, which are tightly coupled with baryons.
\item {\bf Transport sphere}: Beyond the energy sphere, elastic scattering by nucleons and nuclei determines  the opacity. As the rest mass energy of these particles is much larger than the typical neutrino energy, scattering can be treated as elastic.\footnote{Note that Ref.~\citep{raff01} investigated how the recoil term affects the neutrino spectrum.}
\end{itemize}

For $\nu_\e$ and $\bar\nu_\e$ all the neutrinospheres 
are almost coincident, and thus the spectrum is almost thermal.\footnote{More precisely, because of the energy dependencies  of neutrino-matter interactions high-energy neutrinos decouple further outside at lower matter temperature so that the spectrum becomes pinched from the pure Fermi-Dirac spectrum due to the negative temperature gradient \citep{jank89,tamb12}. } For $\nu_\mu, \nu_\tau,\bar\nu_\mu$, and $\bar\nu_\tau$, on the other hand, the radii of the neutrinospheres are separated from each other \citep{raff01}, and thus neutrinos may establish a  non-thermal component, which would be produced between the energy and transport spheres. This is the region we consider in this paper.

With the matrix element for neutrino-nucleon scattering given in Appendix~\ref{sec:matrix}, the right-hand side of Eq.~(\ref{eq:boltzmann}) becomes
\begin{align}
&{\sum_{N}}\frac{G_F^2}{\pi} p
 \int d p' p' \frac{d \Omega^{\prime}}{4 \pi} \int \frac{d^{3} \bm{q}}{(2 \pi)^{3}}~g_N(\bm{q})
\Big\{
\left[ ({F}_1^N)^2 + 3({F}_A^N)^2 \right]
+ \left[ ({F}_1^N)^2 - ({F}_A^N)^2 \right] \cos \theta
\nonumber\\
&\mp 4 \frac{p}{m_N} ({F}_1^N + {F}_2^N) {F}_A^N (1 - \cos \theta)
- \frac{p}{m_N} ({F}_1^N)^2 (1 - \cos^2 \theta)
- \frac{p}{m_N} ({F}_A^N)^2 (3 - 4 \cos \theta + \cos^2 \theta)
\nonumber\\ 
&-4 [({F}_1^N)^2 + ({F}_A^N)^2] \frac{{\hat p} \cdot {\bf q}}{m_N}
+ [({F}_1^N)^2 - ({F}_A^N)^2] (1 - \cos \theta) 
\frac{{\hat p} \cdot {\bf q} - {\hat p'} \cdot {\bf q}}{m_N}
\Big\}
\nonumber\\
\times&\left\{\delta\left(p-p^{\prime}\right)+\frac{\left(\bm{p}-\bm{p}'\right) \cdot \bm{q}}{m_N} \frac{\partial \delta\left(p-p'\right)}{\partial p'}
+\frac{\left(\bm{p}-\bm{p}'\right)^{2}}{2 m_N} \frac{\partial \delta\left(p-p'\right)}{\partial p'}+\frac{1}{2}\left[\frac{\left(\bm{p}-\bm{p}' \right) \cdot \bm{q}}{m_N}\right]^{2} \frac{\partial^{2} \delta\left(p-p'\right)}{\partial p^{\prime 2}} \right\}
\nonumber\\
\times&
\left\{f^{(0)}\left(p^{\prime}\right)-f^{(0)}(p)-f^{(0)}\left(p^{\prime}\right)\left(1-f^{(0)}(p)\right) \frac{\left(\bm{p}-\bm{p}^{\prime}\right) \cdot\left(\bm{q}-m_N \bm{v}_N\right)}{m_Nk_BT_N}\right.
\nonumber\\
&\qquad \left. +f^{(0)}\left(p^{\prime}\right)\left(1-f^{(0)}(p)\right)\left(\frac{-\left(\bm{p}-\bm{p}^{\prime}\right)^{2}}{2 m_Nk_B T_N}+\frac{1}{2}\left(\frac{\left(\bm{p}-\bm{p}^{\prime}\right) \cdot\left(\bm{q}-m_N \bm{v}_N\right)}{m_N k_B T_N}\right)^{2}\right)  \right\}\,,
\label{eq:expanded}
\end{align}
where $\mp$ takes $-$ for neutrinos and $+$ for  anti-neutrinos, and $f^{(0)}(p)$ is the zeroth-order distribution function of neutrinos, which is a Fermi-Dirac distribution with zero chemical potential. We shall ignore the feedback of distorted neutrino spectrum in this paper ($f^{(1)}$ and $f^{(2)}$ in Ref. \citep{dode95}). Here,  $\hat{p}$ is a unit vector,  $\cos\theta=\hat{p}\cdot\hat{p}'$, $G_F$ is the Fermi coupling constant and $F_1^N$, $F_2^N$ and $F_A^N$ are the form factors appearing in $|M_N|^2$ (see Appendix~\ref{sec:matrix}). The derivative of the Dirac delta function will be handled by integration by parts.

Performing integration over $\bm{q}$ yields
\begin{align}
&\int \frac{d^3q}{(2\pi)^3}~g_N(\bm{q})=n_N\,,\quad
\int \frac{d^3q}{(2\pi)^3}~g_N(\bm{q})\bm{q}=n_Nm_N\bm{v}_N\,,\\
&\int \frac{d^3q}{(2\pi)^3}~g_N(\bm{q})q_iq_j=n_N\left(m_Nk_BT_N\delta_{ij}+m_N^2v_{Ni}v_{Nj}\right)\,.
\end{align}

To perform integration over the solid angle of $\hat p'$, $\int d\Omega'$, we write $\bm{p}'\cdot\bm{v}_N=p'v_N\mu'$ and $\int d\Omega'=\int_0^{2\pi}d\phi'\int_{-1}^1 d\mu'$. Using the addition theorem of Legendre polynomials $P_\ell(x)$, we have $(2\pi)^{-1}\int d\phi'~P_\ell(\hat p\cdot\hat p')=P_\ell(\mu)P_\ell(\mu')$ \citep{dode95}. Therefore, integration over $\phi'$ yields the following substitutions: $\cos\theta\to \mu\mu'$ and $\cos^2\theta\to (1-\mu^2-\mu'{}^2+3\mu^2\mu'{}^2)/2$. Integrating over $\mu'$, we  obtain
\begin{align}
&{\sum_{N}}\frac{G_F^2n_N}{\pi m_N}
\Big\{-m_N v_N  p^3\frac{2}{3}[({F}_1^N)^2 + 5({F}_A^N)^2]  \mu \frac{\partial f^{(0)}}{\partial p}
\nonumber\\
&+  \frac{2}{3}[({F}_1^N)^2 + 5({F}_A^N)^2] \frac1{p^2} 
\frac{\partial}{\partial p} \left[ p^6 \left(k_BT_N \frac{\partial f^{(0)}}{\partial p} + f^{(0)}(p)(1-f^{(0)}(p)) \right) \right]
\nonumber\\
&+ p^4v_N\frac{2}{3}[({F}_1^N)^2 + 7({F}_A^N)^2 \pm 8{F}_A^N({F}_1^N+{F}_2^N)]  \mu \frac{\partial f^{(0)}}{\partial p}
\nonumber\\
&+ p^3 m_N v_N^2 
\left( \frac{2}{3}[({F}_1^N)^2 + 5({F}_A^N)^2](1+3\mu^2) 
\frac{\partial f^{(0)}}{\partial p}
  \right.\nonumber\\ 
&\qquad\qquad\qquad +\left.\frac{p}{6}[({F}_1^N)^2(1+\mu^2) + ({F}_A^N)^2(3+11\mu^2) ] 
\frac{\partial^2 f^{(0)}}{\partial p^2} \right)\Big\}\,.
\end{align}

Finally, we average over the directions of the bulk velocities of the nucleons by performing $\int_{-1}^1 d\mu/2$, and obtain
\begin{align}
\frac{df}{dt}=
\sum_N n_N\sigma_N\left(1-\frac13\delta_N\right)
\left\{
\frac{v_N^2}3\frac1{p^4}
\frac{\partial}{\partial p} \left[ p^6 \frac{\partial f^{(0)}}{\partial p}\right]
+
\frac1{m_Np^4}
\frac{\partial}{\partial p} \left[ p^6 \left(k_BT_N\frac{\partial f^{(0)}}{\partial p}
+f^{(0)}(1-f^{(0)})\right)
\right]
\right\}\,,
\label{eq:final}
\end{align}
where $\sigma_N$ is the total scattering cross section given by 
\begin{eqnarray}
    \sigma_N=\frac{G_F^2}{\pi}p^2\left[({F}_1^N)^2 + 3({F}_A^N)^2\right]=\left(\frac{p}{m_\e}\right)^2\times
\begin{cases}
5.35\times 10^{-45}~{\rm cm^2}~~(\mathrm{for}~N=\p)\\
6.45\times 10^{-45}~{\rm cm^2}~~(\mathrm{for}~N=\n)
\end{cases}
\label{eq:sigma_N}
\end{eqnarray}
and the symbol \citep{burr06b}
\begin{align}
\delta_N \equiv \frac{({F}_1^N)^2 - ({F}_A^N)^2}{({F}_1^N)^2 + 3({F}_A^N)^2}=
\begin{cases}
-0.33~~(\mathrm{for}~N=\p)\\
-0.11~~(\mathrm{for}~N=\n)
\end{cases}
\label{eq:delta_N}
\end{align}
captures angular dependence of scattering in the rest frame of nucleons, $d\sigma_N/d\Omega=(\sigma_N/4\pi)(1+\delta_N\cos\theta)$ ---see the first line of Eq.~(\ref{eq:expanded}).

Eq.~(\ref{eq:final}) is the main result of this paper. Comparing this to the result for photon-electron scattering (see Eq.~(15) of \cite{hu94}), 
\begin{align}
\frac{df}{dt}=n_\e\sigma_T
\left\{
\frac{v^2_\e}3\frac1{p^2}
\frac{\partial}{\partial p} \left[ p^4 \frac{\partial f^{(0)}}{\partial p}\right]
+
\frac1{m_\e p^2}
\frac{\partial}{\partial p} \left[ p^4 \left(k_BT_\e\frac{\partial f^{(0)}}{\partial p}
+f^{(0)}(1+f^{(0)})\right)
\right]
\right\}\,,
\label{eq:photon}
\end{align}
we find two differences. First, the power of $p$ is different by two because the neutrino-nucleon cross section is proportional to $p^2$ whereas the Thomson scattering cross section $\sigma_T$ is independent of the photon momenta. Second, we have $1-f^{(0)}$ in the last term in Eq.~(\ref{eq:final}) instead of $1+f^{(0)}$ in Eq.~(\ref{eq:photon}), because of Fermi statistics.

\section{Solutions}
\label{sec:solution}
\subsection{Chemical potential distortion}
From now on we shall assume that protons and neutrons share the same temperature $T$ and bulk velocity $V$. We shall use the approximation that the masses of protons and neutrons are the same and denote them as $m$, i.e., $m_N\to m$, and drop the superscript $(0)$ on $f$.

Let us define dimensionless variables $x\equiv p/(k_BT)$ and 
\begin{align}
\nonumber
    dy&\equiv \sum_N\sigma_N(x=1) \left(1-\frac13\delta_N\right)n_N\frac{k_BT}{m}dt\\
    &=\left<n \sigma(x=1)\right>\frac{k_BT}{m_N}dt,
\end{align}
where
\begin{align}
    \left<n \sigma\right>=\sum_Nn_N\sigma_N\left(1-\frac{1}{3}\delta_N\right)
    =n_\n\left(1-\frac{1}{3}\delta_\n\right)\sigma_\n+n_\p\left(1-\frac{1}{3}\delta_\p\right)\sigma_\p.
    \label{eq:nsigma}
\end{align}
We then write Eq.~(\ref{eq:final}) as
\begin{equation}
    \frac{df}{dy}=\frac1{x^2}\frac{\partial}{\partial x}\left\{x^6\left[f-f^2+(1+\delta_v)f'\right]    \right\}\,,
\label{eq:kompaneets}
\end{equation}
where the prime denotes derivative with respect to $x$, and $\delta_v\equiv mV^2/(3k_BT)$ is the ratio of nucleon's bulk kinetic energy to thermal energy. 

An equilibrium solution, $df/dy=0$, is given by
\begin{equation}
f(p)=\frac{1}{e^{(p-\mu_\nu)/[k_BT(1+\delta_v)]}+1}\,,
\label{eq:spectrum}
\end{equation}
where $\mu_\nu$ is a chemical potential.  The bulk velocity
effectively enlarges the temperature by a factor of $1+\delta_v$.

As the number of neutrinos is conserved in our setup, we can obtain the emergent spectrum after stochastic scattering by bulk fluid
motion as follows. For an initially thermal spectrum (i.e. $\mu_\nu=0$), 
we get the same number density of neutrinos if we have $\mu_\nu=-0.341$, $-4.34$, and $-80.3k_BT$ for
$\delta_v=0.1$, $1$, and $10$, respectively. 
Therefore, scattering distorts the neutrino thermal spectrum by giving a non-zero chemical potential, which is well known as ``$\mu$-distortion'' in the cosmic microwave background research \citep{zeld69}.

With the spectrum of Eq. (\ref{eq:spectrum}), we can calculate the
neutrino-annihilation rate which is one of the important heating
processes in supernova explosions. The energy deposition rate via neutrino annihilation
($\nu+\bar\nu\to \mathrm{e}^++\mathrm{e}^-$) is given by
\citep{good87,seti06}
 \begin{equation}
 \dot E_{\nu\bar\nu}=\mathcal{C}F_{3,\nu}F_{3,\bar\nu}
 \left(\frac{\bracket{\epsilon_\nu^2}\bracket{\epsilon_{\bar\nu}}+\bracket{\epsilon_{\bar\nu}^2}\bracket{\epsilon_{\nu}}}{\bracket{\epsilon_{\nu}}\bracket{\epsilon_{\bar\nu}}}\right)\,,
 \end{equation}
where $F_{i,\nu}\equiv\int p^idpf$,
$\bracket{\epsilon_\nu}=F_{3,\nu}/F_{2,\nu}$ and
$\bracket{\epsilon_\nu^2}=F_{4,\nu}/F_{2,\nu}$. The subscript $\bar\nu$ indicates the corresponding quantities for anti-neutrinos. 
The overall factor $\mathcal{C}$ includes the weak interaction coefficients and
information on the angular distribution of the neutrinos. To
calculate this factor we need to determine the geometry of the
neutrino-emitting source. Since this factor is expected not to change
significantly by including the neutrino acceleration process, we
concentrate on the effect of the spectral change here. For simplicity,
we assume that the spectra of $\nu$ and $\bar\nu$ are identical. Then,
we obtain
\begin{equation}
\dot E_{\nu\bar\nu}\propto \frac{F_{3,\nu}^2\bracket{\epsilon_\nu^2}}{\bracket{\epsilon_\nu}}
\propto \bracket{\epsilon_\nu}\bracket{\epsilon_\nu^2}\,.
\end{equation}
Here we use $F_{3,\nu}\propto \bracket{\epsilon_\nu}$, as
$F_{2,\nu}$ does not change by scattering processes alone. Roughly speaking,
$\bracket{\epsilon_\nu}=(1+\delta_v)\bracket{\epsilon_\nu}_\mathrm{thermal}$
and
$\bracket{\epsilon_\nu^2}=(1+\delta_v)^2\bracket{\epsilon_\nu^2}_\mathrm{thermal}$,
where $\bracket{\epsilon_\nu}_\mathrm{thermal}$ and
$\bracket{\epsilon_\nu^2}_\mathrm{thermal}$ are mean energy and mean
square energy based on a purely thermal distribution
function. Therefore, we find
\begin{equation}
\dot E_{\nu\bar\nu}\propto (1+\delta_v)^3\,.
\end{equation}

\subsection{Including opacity}
\label{sec:extended_boltzmann}

The calculation of the energy deposition rate given above includes only scattering; thus, it can describe only $\mu$- and $\tau$-type neutrinos.
However, there are also terms related to emission and absorption. Absorption is important for electron-type neutrinos and leads to thermalization of the spectrum of $\nu_e$ and $\bar\nu_e$. In the following we include all relevant terms. 

The revised transfer equation in which both ${\mathcal O}(v)$ and ${\mathcal
  O}(v^2)$ terms are taken into account is written as follows
(see also Refs. \cite{psal97,tita97} for photon cases):
\begin{align}
\frac{df}{dt}&=
\nabla\cdot\left(\frac{1}{3\kappa^\mathrm{t}}\nabla f\right)
+\frac{1}{3}\left(\nabla\cdot{\bm V}\right)p\frac{\partial f}{\partial p}+\kappa^\mathrm{a}(f^{\mathrm{eq}}-f)\nonumber\\
&+
\frac1{p^2}\frac{\partial}{\partial p}
\left\{\frac{\kappa^\mathrm{sc}}{m}p^4\left[f-f^2+\left(k_BT+\frac{mV^2}{3}\right)\frac{\partial f}{\partial p}\right]\right\}\,,
\label{eq:transfer}
\end{align}
where $\kappa^\mathrm{a}$ and $\kappa^\mathrm{sc}$
are the opacities (inverse of the mean free path) for absorption and
scattering, respectively, $\kappa^\mathrm{t}\equiv\kappa^\mathrm{a}+\kappa^\mathrm{sc}$,
and $f^\mathrm{eq}$ is the neutrino distribution function in thermal
equilibrium. When we omit the second line on the right-hand side of this equation, we get  exactly the same as the equation for neutrino transfer based on
diffusion approximation, which is solved in a number of numerical
simulations; see Eq. (A27) of Ref. \citep{brue85} for a flux-limited diffusion
approximation and Eqs. (5) and (6) of Ref. \citep{lieb09} for an isotropic
diffusion source approximation. Both of them were originally derived
from the Boltzmann equation including velocity-dependent terms up to
${\mathcal O}(v)$ and are commonly used in neutrino-radiation
hydrodynamics simulations. 

The effect of the second term proportional to $\nabla\cdot{\bm
  V}$ on the right-hand side is analogous to the bulk
Comptonization process in photon cases and the first-order Fermi
acceleration for charged particles. 
This term also modifies the neutrino
spectrum when there is a compressional flow, such as a shock
or an accretion flow onto a black hole. In core-collapse supernovae, a neutron star is formed and compression is almost negligible in the optically thick region for neutrinos so that this first-order
acceleration does not work at all \citep{suwa13a}.

\subsection{Numerical solution}
\label{sec:numerical_solution}
Assuming one-zone (i.e. $\nabla f=0$) and incompressible bulk flow
(i.e. $\nabla\cdot{\bm V}=0$), Eq.~(\ref{eq:transfer}) reads
\begin{align}
\frac{df}{dt}=\kappa^\mathrm{a}(f^{\mathrm{eq}}-f)+
\frac{\partial}{\partial p}
\left\{\frac{\kappa^\mathrm{sc}}{m}p^4\left[f-f^2+\left(k_BT+\frac{mV^2}{3}\right)\frac{\partial f}{\partial p}\right]\right\}\,.
\label{eq:small_bulk_vel}
\end{align}
The scattering opacity is given from Eqs. \eqref{eq:sigma_N}, \eqref{eq:delta_N} and \eqref{eq:nsigma} as
\begin{align}
\nonumber
    \kappa^\mathrm{sc}=\left<\sigma n\right>
    &=\left[\sigma_\n\left(1-\frac{1}{3}\delta_\n\right)Y_\n+\sigma_\p\left(1-\frac{1}{3}\delta_\p\right)Y_\p\right]\frac{\rho}{m}\\
    &= 4.0\times 10^{-10}~\mathrm{cm}^{-1}\left(\frac{p}{m_\mathrm{e}}\right)^2\left(\frac{\rho}{10^{11}~\mathrm{g~cm}^{-3}}\right)\left(1-0.11Y_\p\right),
\end{align}
where $Y_\n=n_\n/n_\mathrm{b}$ and
$Y_\p=n_\p/n_\mathrm{b}$ are the number fractions of free neutrons and protons with $n_\mathrm{b}$ being the number density of baryons, and $m=1.67\times 10^{-24}$~g. Since all nuclei are photodisintegrated into neutrons and protons, $Y_\n+Y_\p=1$ here. The absorption opacity ($\nu_\e+\n\to \p+\e^-$ for $\nu_\e$ and $\bar\nu_\e+\p\to \n+\e^+$ for $\bar\nu_\e$) is given by \cite{burr06b}
\begin{align}
\kappa^\mathrm{a}&=\frac{3g_A^2+1}{4}\sigma_0\left(\frac{p}{m_\mathrm{e}}\right)^2\frac{\rho}{m}\times
\begin{cases}
Y_\n~~(\mathrm{for}~\nu_\mathrm{e})\\
Y_\p~~(\mathrm{for}~\bar\nu_\mathrm{e})
\end{cases}\nonumber\\
&= 7.5\times 10^{-10}~\mathrm{cm}^{-1}\left(\frac{p}{m_\mathrm{e}}\right)^2\left(\frac{\rho}{10^{11}~\mathrm{g~cm}^{-3}}\right)\left(\frac{Y}{0.5}\right)\,,
\label{eq:kappa^a}
\end{align}
where $g_A=1.2723$ is the axial-vector coupling constant, and $\sigma_0=1.705\times 10^{-44}$ cm$^2$ is a reference neutrino cross
section. The small corrections due to the mass difference between a
neutron and a proton, and to weak magnetism and recoil are neglected here. For characteristic values
 $Y\equiv
Y_\n=Y_\p=0.5$ is used in Eq. (\ref{eq:kappa^a}).

For constant temperature and chemical potential, $V^2=0$,
and $f^\mathrm{eq}_\nu=(e^{(\epsilon_\nu-\mu_\nu)/k_BT}+1)^{-1}$, we
have a steady-state solution ($\partial f/\partial t=0$) of
Eq. (\ref{eq:small_bulk_vel}) given by $f=f^\mathrm{eq}_\nu$. This solution is the same as the steady-state solution of the equation without the second term on the right-hand side in Eq.~(\ref{eq:small_bulk_vel}).

Writing the Boltzmann equation in a dimensionless form, we have
\begin{align}
\frac{\partial f}{\partial y}=
\frac{1}{x^2}\frac{\partial}{\partial x}
\left\{
  x^6
  \left[
    f-f^2+(1+\delta_v)f'
  \right]
\right\}
+\Theta_T^{-1}(f^\mathrm{eq}-f)\,,
\label{eq:nondimensional}
\end{align}
where
$\Theta_T\equiv
(\kappa^\mathrm{sc}/\kappa^\mathrm{a})
(k_BT/m)$. 

\begin{figure}
\includegraphics[width=1\textwidth]{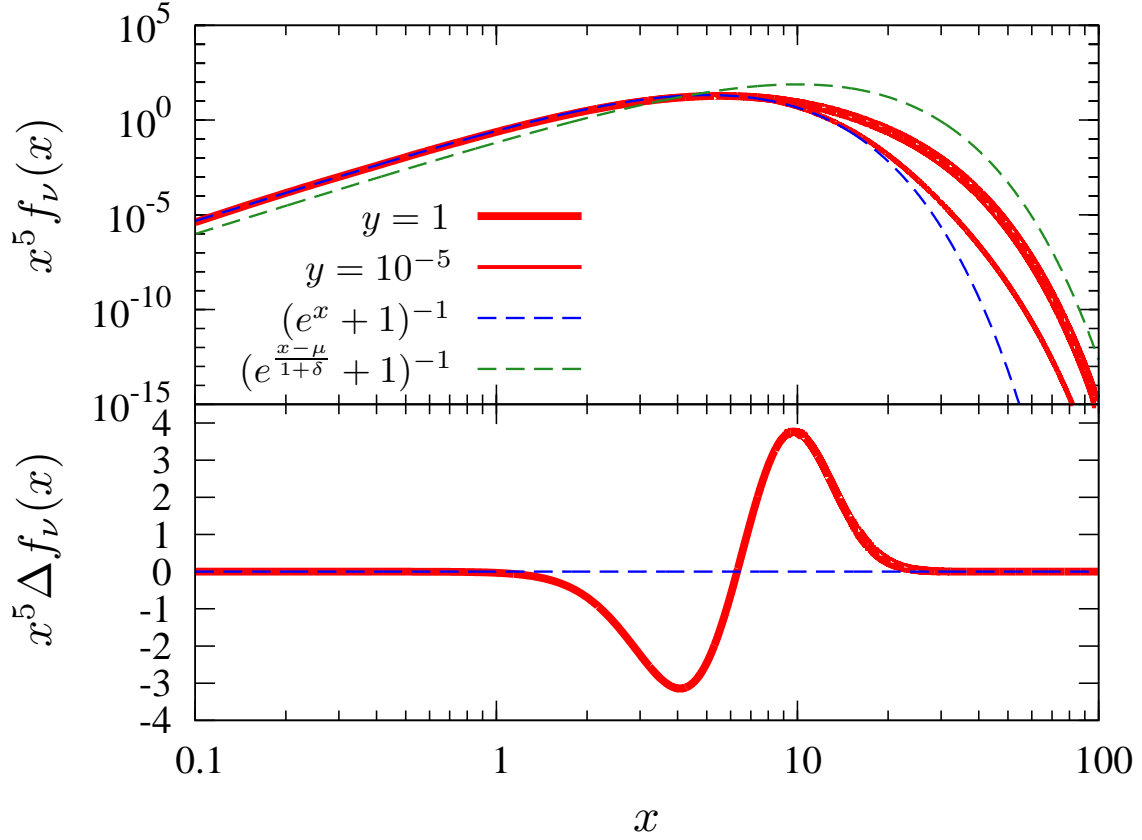}
\caption{{\it (Top)} Numerical solutions for $k_BT=6$ MeV,
  $\delta_v=1$, and $Y_\p=0.1$ (red lines). 
  The initial spectrum and an analytic solution without absorption
  are shown by the blue- and green-dashed lines, respectively. For numerical solutions, two different values of $y$ are
  used ($y=1$ and $10^{-5}$ for the thick and thin lines, respectively).  {\it
    (Bottom)} Difference between numerical solutions at $y=1$ and
  the initial spectrum. The low-energy part ($x\lesssim 7$) is reduced, while
  the high-energy part is increased due to up-scattering.}
\label{fig:spec}
\end{figure}

Figure \ref{fig:spec} shows numerical solutions of
Eq. (\ref{eq:nondimensional}) as red solid lines.  For
this calculation we use $k_BT=6$ MeV, $\delta_v=1$, and
$Y_\p=0.1$. Since we use $\kappa^\mathrm{a}\propto Y_\p$,
the solutions shown in this figure are valid for $\bar\nu_e$. The initial
condition at $y=0$ is given by $(e^x+1)^{-1}$, which is shown by the 
blue-dashed line in this figure. The green-dashed line is an analytic
solution of Eq.~(\ref{eq:spectrum}), which is realized when the
absorption is not taken into account. We find that, for higher
energy ($x\gtrsim 10$), the numerical solutions deviate from the thermal equilibrium (blue-dashed line), and the spectra become slightly non-thermal toward the analytic solution (green-dashed line). This is because the ratio of the two terms on the right-hand side of
Eq.~(\ref{eq:small_bulk_vel}) is $\propto p/m$, such that the absorption term dominates in the low-energy regime.
We find that the numerical solutions are very similar for $y\gtrsim 0.001$.

\subsection{Implications for neutrino heating rate}

Since the neutrino interaction rate is a strong function of neutrino
energy, the neutrino heating rate is affected when the neutrino
spectrum changes. The neutrino heating rate is given as \citep{jank01}
\begin{equation}
Q_\nu^+\propto E_\nu^2L_\nu\,,
\label{eq:qnu}
\end{equation}
where 
\begin{align}
&E_\nu^2\propto \frac{\int^\infty_0  dp~p^5 f}{\int^\infty_0  dp~p^3 f}\,,
\label{eq:ebra}\\
&L_\nu\propto \int^\infty_0 dp~p^3 f\,.
\label{eq:Lnu}
\end{align}
With the numerical solutions for the neutrino spectrum (Figure
\ref{fig:spec}), we find the following results:
\begin{align}
\frac{\int_0^\infty f(x)x^5dx}{\int_0^\infty f^\mathrm{eq}(x)x^5dx}
&=1+0.017~\delta_v \left(\frac{k_BT}{1~\mathrm{MeV}}\right)=1+0.053\left(\frac{V}{0.1}\right)^2\,,
\label{eq:f5}\\
\frac{\int_0^\infty f(x)x^3dx}{\int_0^\infty f^\mathrm{eq}(x)x^3dx}
&=1-0.010~\delta_v \left(\frac{k_BT}{1~\mathrm{MeV}}\right)=1-0.031\left(\frac{V}{0.1}\right)^2\,.
\label{eq:f3}
\end{align}
They are valid for $\delta_v\lesssim 1$. For larger $\delta_v$, non-linear terms in $\delta_v$ contribute. Combining them gives an enhancement in the square mean energy over its thermal value as
\begin{equation}
\frac{E_\nu^2}{E_{\nu,\mathrm{thermal}}^2}=1+0.086\left(\frac{V}{0.1}\right)^2\,.
\end{equation}
Thus, the neutrino heating rate $Q_\nu^+$ is amplified inside the gain region, where the neutrino heating overwhelms neutrino cooling between the shock and the neutrinospheres
---see Eqs. \eqref{eq:qnu} and \eqref{eq:f5}. The neutrino cooling rate is reduced inside the protoneutron star ---see
Eqs. \eqref{eq:Lnu} and \eqref{eq:f3}. Both of them make shock revival
of a supernova easier. The actual velocity would be $V\sim$ 0.02--0.08 for the turbulent motion. To give an accurate value of the correction, we need hydrodynamical simulations.

\section{Summary and discussion}
\label{sec:summary}
We have derived a Kompaneets-like equation for neutrinos by expanding the collision integral of the Boltzmann equation up to the second order in  energy transfer, or ${\mathcal O}(v^2)$, including thermal and bulk velocities of nucleons. We also included absorption and emission of neutrinos, to arrive at the full neutrino transport equation given in  Eq. (\ref{eq:transfer}). The dimensionless form of the  equation suitable for numerical calculations is given in Eq.~(\ref{eq:nondimensional}), and the numerical solutions are presented in Figure \ref{fig:spec}. 

We find that the distortion of the neutrino spectrum due to interaction with nucleons leads to a larger neutrino heating rate in
the gain region and a smaller neutrino cooling rate in the protoneutron star, which provides a better condition for supernova explosions than solutions without the effects we find in this paper.

The formulation given in this paper is a natural extension of the transfer equation that is solved in some neutrino-radiation hydrodynamics
simulations for core-collapse supernovae, in which velocity-dependent
terms in the collision integral are included only up to ${\mathcal O}(v)$. 
However, this effect would be implicitly included in numerical simulations if the transfer equation is solved in a comoving frame and the Lorentz transformation is properly performed between the comoving and laboratory frames.
Note
that the current equation is derived in the non-relativistic limit,
i.e. $k_BT\ll m$ and $V\ll 1$. Relativistic corrections are known to amplify the
spectral distortion of the SZ effect (e.g. see Ref. \citep{noza00}), which could
even enhance the neutrino heating rates.

The numerical solutions presented in Section
\ref{sec:numerical_solution} give the emergent spectrum of the
electron-type neutrino ($\nu_e$) and anti-neutrinos ($\bar\nu_e$), as we include the absorption and
emission term (the last term on the right-hand side in
Eq.~\ref{eq:nondimensional}), which is relevant only for
charged-current reactions. For other heavier leptonic flavors,
i.e. muon-type and tauon-type, this term does not appear in the
transfer equation so that their spectrum would be like
Eq.~(\ref{eq:spectrum}). The neutrino annihilation rate of these
heavier leptonic neutrinos \citep{good87,jank91} can also be amplified
by a factor $(1+\delta_v)^3$, which would make supernova explosion
easier as well.

Our finding motivates self-consistent hydrodynamics simulations including the spectral distortion of neutrinos, which will be needed to  calculate the quantitative impact on supernova explosions. Our solutions would also be useful in testing the numerical code handling the capability of the terms of $\mathcal{O}(v^2)$.

\section*{Acknowledgments}
E.~K. thanks Jens Chluba for useful discussions.
Y.~S. and H.~W.~H.~T. thank the Max Planck Institute for Astrophysics, where this work was initiated and completed, for their hospitality. Y.~S. was supported in part by Japan Society for the Promotion of Science (JSPS) postdoctoral fellowships for research abroad. This study was supported in part by the Grant-in-Aid for Scientific Research (Nos. 15H05896, 16H00869, 16K17665, 17H02864, 18H04586, and 18H05437). H.~W.~H.~T. was supported in part by the Advanced Leading Graduate Course for Photon Science (ALPS).

\appendix
\section{Neutrino-nucleon scattering}
\label{sec:matrix}
\subsection{Matrix element}
In this appendix, we follow Refs. \citep{smith71,leitner05} to write an expression for the spin-averaged matrix element of neutrino-nucleon scattering. It is given by
\begin{align}
|M_N|^2 = 16 G_F^2 m_N^4
\left[A \mp \frac{s-u}{m_N^{2}} B+\frac{(s-u)^{2}}{m_N^{4}} C\right]\,,
\end{align}
where $\mp$ takes $-$ for neutrinos and $+$ for anti-neutrinos, and 
\begin{align}
s&=(p+q)^2= m_N^2 + 2 p \cdot q \,,\quad
u=(p'-q)^2= m_N^2 - 2 p' \cdot q\,,\quad
\tau=\frac{Q^{2}}{4 m_N^{2}}\,,
\\
A&=4\tau\left[(1+\tau)\left({F}_{A}^{N}\right)^{2}-(1-\tau)\left({F}_{1}^{N}\right)^{2}+\tau(1-\tau)\left({F}_{2}^{N}\right)^{2}+4 \tau {F}_{1}^{N} {F}_{2}^{N}\right]\,,
\\
B&=4\tau {F}_{A}^{N}\left({F}_{1}^{N}+{F}_{2}^{N}\right)\,,\quad
C=\frac{1}{4}\left[\left({F}_{A}^{N}\right)^{2}+\left({F}_{1}^{N}\right)^{2}+\tau\left({F}_{2}^{N}\right)^{2}\right]\,,
\end{align}
where all momenta are four-vectors, and ${F}_1^N$, ${F}_2^N$ and ${F}_A^N$ are the so-called neutral-current Dirac, Pauli and axial form factors, respectively, which depend on $Q^2\equiv -(p-p')^2$.

When nucleons are non-relativistic, we have
\begin{align}
p \cdot q &= p \; m_N - {\bf p} \cdot {\bf q} 
= p_{c.m.} m_N + p_{c.m.}^2\,,
\\
p' \cdot q &= p' m_N - {\bf p'} \cdot {\bf q} 
= p_{c.m.} m_N + p_{c.m.}^2 \cos{\theta_{c.m.}}\,,
\\
p \cdot p' &= p p' (1 - {\hat p} \cdot {\hat p'} ) 
= p_{c.m.}^2 (1-\cos{\theta_{c.m.}})\,,
\end{align}
where the momenta on the left-hand side are four-vectors, whereas those in the middle and on the right-hand side are three-vectors and their magnitudes. 
$p_{c.m.}$ and $\theta_{c.m.}$ are the energy and scattering angle in the center-of-mass frame. Eliminating $p_{c.m.}$ and $\theta_{c.m.}$, we obtain
\begin{align}
p'= p \frac{m_N-{\hat p} \cdot {\bf q}}{m_N-{\hat p'} \cdot {\bf q} + p (1 - {\hat p} \cdot {\hat p'} )}\,. 
\end{align}
Using this relationship to expand the matrix element in powers of $p/m_N$ and ${\bf q}/m_N$ up to their linear order, we obtain
\begin{align}
|M_N|^2 &= 16 G_F^2 m_N^2 p^2 
\left\{
\left[ ({F}_1^N)^2 + 3({F}_A^N)^2 \right]
+ \left[ ({F}_1^N)^2 - ({F}_A^N)^2 \right] \cos \theta
\right. \nonumber\\
&\mp 4 \frac{p}{m_N} ({F}_1^N + {F}_2^N) {F}_A^N (1 - \cos \theta)
- \frac{p}{m_N} ({F}_1^N)^2 (1 - \cos^2 \theta)
- \frac{p}{m_N} ({F}_A^N)^2 (3 - 4 \cos \theta + \cos^2 \theta)
\nonumber\\ 
&\left. -4 [({F}_1^N)^2 + ({F}_A^N)^2] {\hat p} \cdot \frac{\bf q}{m_N}
+ [({F}_1^N)^2 - ({F}_A^N)^2] (1 - \cos \theta) 
\frac{{\hat p} \cdot {\bf q} - {\hat p'} \cdot {\bf q}}{m_N}
\right\}\,,
\end{align}
where ${\hat p} \cdot {\hat p'} \equiv \cos \theta$.

\subsection{Total cross section}
In the center-of-mass frame, the differential cross section is given by 
\begin{align}
\left. \frac{d\sigma_N}{d\Omega} \right|_\text{c.m.}
=
\frac{1}{64\pi^2 s} |M_N|^{2}\,.
\end{align}
Integrating this over angles, we obtain the total cross section in the limit of non-relativistic nucleons ($s\approx m_N^2$) as
\begin{align}
\sigma_N
=
\frac{G_F^2}{\pi} p^2 \left[ ({F}_1^N)^2 + 3({F}_A^N)^2 \right]\,.
\end{align}

For the energy scale we are interested in, the only relevant form factors are the ones with $Q^2=0$. Thus, 
\begin{align}
2 {F}_{1}^{\p}(0)&
=1-4 \sin ^{2} \theta_{W}\,,\quad
2 {F}_{1}^{\n}(0)=-1\,,\quad 2 {F}_{A}^{\p}(0)=g_A\,,\quad
2 {F}_{A}^{\n}(0)=-g_A\,,
\\
2 {F}_{2}^{\p}(0)
&=\left(1-4 \sin ^{2} \theta_{W}\right)\left( \mu_\p - 1 \right)-\mu_\n\,,\quad
2 {F}_{2}^{\n}(0)=\left(1-4 \sin ^{2} \theta_{W}\right)\mu_\n-\left(\mu_\p-1\right)\,.
\end{align}
Here, $g_A=1.2723$ is the axial-vector coupling constant, $\sin^2\theta_W=0.23122$ is the weak angle, $G_F=1.1663787\times 10^{-5}~{\rm GeV}^{-2}$ is the Fermi coupling constant, and $\mu_\p$ and $\mu_\n$ are the magnetic moments of protons and neutrons, respectively. As ${F}_{2}^{N}$ does not contribute to the angle-averaged collision integral in Eq. \eqref{eq:final}, we do not need to evaluate them here. We thus find that
\begin{align}
\sigma_\p
&=
\frac{G_F^2}{4\pi} p^2 \left[ (1-4\sin^2\theta_W)^2 + 3g_A^2 \right]=
5.35\times 10^{-45}~{\rm cm^2}\cdot \left( \frac{p}{m_\e} \right)^2\,,
\\
\sigma_\n
&=
\frac{G_F^2}{4\pi} p^2 \left[ 1 + 3g_A^2 \right]=
6.45\times 10^{-45}~{\rm cm^2} \cdot \left( \frac{p}{m_\e} \right)^2\,.
\end{align}
 
\section{Alternative derivation \`a la Rybicki and Lightman}
\label{sec:rybicki}
In this appendix, we provide an alternative derivation of the main result of this paper, Eq.~(\ref{eq:final}), following the argument given in Section 7.6 of Ref. \citep{rybi79}. For ease of comparison, let us change our notation of momenta to follow closely that of Ref. \citep{rybi79}: $\bm{q}\to \bm{p}$, $\bm{q}'\to\bm{p}_1$, $\bm{p}\to \omega\hat{n}$, and $\bm{p}'\to \omega_1\hat{n}_1$.

The Boltzmann equation is
\begin{align}
{\frac{d f(\omega,\hat n,t)}{cd t}}=
&
{\sum_{N}}\int d^3p\int d\Omega~\frac{d\sigma_N}{d\Omega}&\nonumber\\
&\times\{
g_N({\bm p_1})f(\omega_1,\hat n_1,t)[1-f(\omega,\hat n,t)]-g_N({\bm p})f(\omega,\hat n,t)[1-f(\omega_1,\hat n_1,t)]
\}\,,
\label{eq:boltzmann_eq}
\end{align}
where $d\sigma_N/d\Omega=(\sigma_N/4\pi)(1+\delta_N\cos\theta)$ (with
$N=\p$ and $\n$) is the differential cross section of scattering by free nucleons, $g_N({\bm p})$ the distribution function of nucleons with
three-momenta ${\bm p}$, $\omega$ and $\omega_1$  the neutrino
energies,  $\hat n$ and $\hat n_1$ the unit vectors of the neutrino
three-momenta, and $\cos\theta=\hat n\cdot\hat n_1$.

\subsection{Thermal nucleons}
Since the typical energy transfer is small compared to a nucleon's
kinetic energy, we define
\begin{equation}
\Delta\equiv \frac{\omega_1-\omega}{k_BT}\ll 1\,,
\label{eq:delta}
\end{equation}
and expand the right-hand side of Eq.~(\ref{eq:boltzmann_eq}) up to the second order in
$\Delta$. 

In a frame in which the initial nucleon velocity is
$\boldsymbol\beta={\bm p}/E$ with $E=|{\bm p}|^2/(2m_N)$, we have
\begin{align}
\nonumber
\frac{\omega_1}{\omega}
&=\frac{1-\boldsymbol\beta\cdot \hat n}{1-\boldsymbol\beta\cdot \hat
 n_1+\omega(1-\hat n\cdot\hat n_1)
{\sqrt{1-\beta^2}/m_N}}\\ 
&= 1+\frac{{\bm p}\cdot(\hat n_1-\hat n)}{m_N}+\mathcal{O}\left(\left(\frac{k_BT}{m_N}\right)^2\right)\,.
\label{eq:e1/e}
\end{align}
Note that $\beta\equiv|\boldsymbol\beta|\approx
k_BT/m_N\ll 1$. Comparing Eqs. (\ref{eq:e1/e}) and (\ref{eq:delta}), we find
\begin{equation}
\Delta=x(\hat n_1-\hat n)\cdot\frac{{\bm p}}{m_N},
\end{equation}
where $x\equiv \omega/(k_BT)$.

%
Expanding the neutrino distribution function with $\omega_1$ up to
second order in $\Delta$, we obtain
\begin{equation}
f(\omega_1,\hat n_1)\approx f(\omega,\hat n)
+\Delta f'(\omega,\hat n) +\frac{\Delta^2}2 f''(\omega,\hat n)\,,
\end{equation}
where the primes denote derivatives with respect to $\omega$.  We also expand the distribution
function of non-relativistic nucleons up to second order in
$\Delta=(E-E_1)/(k_BT)$ as
$g_N(E_1)\approx\left(1+\Delta+\Delta^2/2\right)g_N(E)$
with
$g_N(E)=n_N(2\pi m_N k_BT)^{-3/2}e^{-E/k_BT}$.
Here, $n_N$ is the number density of nucleons.
Then Eq. (\ref{eq:boltzmann_eq}) becomes
\begin{align}
{\frac{d f}{d t}}&=
\left[f'+f(1-f)\right]\sum_N\int d^3p\int d\Omega\frac{d\sigma_N}{d\Omega}g_N\Delta\nonumber\\
&+\left[\frac{1}{2}f''+\left(f'+\frac{f}{2}\right)(1-f)\right]\sum_N\int d^3p\int d\Omega\frac{d\sigma_N}{d\Omega}g_N\Delta^2\,.
\label{eq:dfdt}
\end{align}
To check the
calculation, note that a Fermi-Dirac distribution
$f_\mathrm{FD}(x)=1/(e^{x+\eta}+1)$ is a steady-state solution to
Eq. (\ref{eq:dfdt}). Here, $\eta$ is an integration constant
corresponding to a chemical potential term.

Let us evaluate the integral in the second term of Eq.~(\ref{eq:dfdt}).  Writing
$(\hat n_1-\hat n)\cdot{\bm p}=|\hat n_1-\hat n|p\cos\xi$ and $\hat
n\cdot\hat n_1=\cos\theta$, we have
\begin{equation}
\Delta^2=2x^2\frac{p^2}{m_N^2}\cos^2\xi (1-\cos\theta)\,.
\end{equation}
Using $d^3p=p^2 dp d(\cos\xi) d\varphi$ and $d\Omega=d(\cos\theta)
d\phi$, the integral evaluates to
\begin{align}
\int d^3p\int d\Omega\frac{d\sigma_N}{d\Omega}g_N\Delta^2
&=n_Nx^2\frac{k_BT}{m_N}\sigma_N\left(2-\frac{2}{3}\delta_N\right)\,.
\end{align}

The integral in the first term of Eq.~(\ref{eq:dfdt}) is more
complicated. Therefore, we follow Ref. \citep{rybi79} and
use the conservation law to calculate it. Since scattering does not change the number of neutrinos, $\int dx x^2\partial f/\partial t=0$.
This equation is satisfied by the continuity equation,
\begin{equation}
{\frac{d f}{d t}}=
-\frac{1}{x^2}\frac{\partial}{\partial x}[x^2 j(x)]\,,
\label{eq:cons}
\end{equation}
where $j$ is the flux of neutrinos in momentum space. 
Next, we write the unknown integral as
\begin{equation}
\int d^3p\int d\Omega\frac{d\sigma_N}{d\Omega}g_N\Delta=n_N\sigma_N\frac{k_BT}{m_N}I\,.
\end{equation}
The Boltzmann equation then becomes
\begin{align}
{\frac{d f}{d t}}=&
\sum_N n_N\sigma_N\frac{k_BT}{m_N}\left[
\alpha_N x^2f''+2\alpha_N x^2f'(1-f)+f'I
+f(1-f)(\alpha_N x^2+I)
\right]\,,
\label{eq:dfdt2}
\end{align}
where $\alpha_N\equiv 1-\delta_N/3$. Requiring that this equation be equal to
Eq.~(\ref{eq:cons}), we write an {\it ansatz},
\begin{equation}
j(x)=g(x)\left[f'+h(f,x)\right]\,,
\end{equation}
with two unknown functions $g$ and $h$, so that $j'$ has a term in
$f''$ but no higher derivatives. We then determine the function
$h$ from the equilibrium form, $f_\mathrm{FD}$, which has the
following relation;
$f_\mathrm{FD}'+f_\mathrm{FD}(1-f_\mathrm{FD})=0$. If $j$ vanishes in
equilibrium then $h=f(1+f)$. We find that
\begin{align}
-\frac{1}{x^2}\frac{\partial}{\partial x}(x^2j)=-
\left\{
gf''+f'
\left[
g'+g\left(1-2f+\frac{2}{x}\right)\right]
+f(1-f)\left(g'+\frac{2g}{x}\right)
\label{eq:deriv_j}
\right\}\,.
\end{align}
Comparing the $f''$ terms in Eqs. (\ref{eq:dfdt2}) and
(\ref{eq:deriv_j}), we obtain
\begin{align}
g=-\sum_Nn_N\sigma_N\frac{k_BT}{m_N}\alpha_N x^2\,.
\end{align}
Comparing the $f(1-f)$ terms gives
\begin{align}
\sum_Nn_N\sigma_N \frac{k_BT}{m_N}(\alpha_N x^2+I)
=-\left(g'+\frac{2g}{x}\right)
=\sum_Nn_N\sigma_N\frac{k_BT}{m_N}\alpha_N(4x+2x)\,,
\end{align}
which yields $I=\alpha_N(6-x)x$.
Note that this makes $f'$ terms consistent as well.

The final result is the following equation:
\begin{equation}
{\frac{d f}{d t}}=
\sum_N
\alpha_N\frac{k_BT}{m_N}\frac{1}{x^2}\frac{\partial}{\partial x}
\left[n_N\sigma_N(x)x^4(f-f^2+f')\right]\,.
\label{eq:kompaneets_dimensional}
\end{equation}
By changing the variable from $t$ to $y$ via $dy\equiv
\sum_N\alpha_N\frac{k_BT}{m_N}n_N\sigma_N(x=1)dt$, we obtain
\begin{equation}
{\frac{d f}{dy}}=
\frac{1}{x^2}\frac{\partial}{\partial x}
\left[
  x^6
  \left(
    f-f^2+f'
  \right)
\right],
\label{eq:kompaneets-wo-v}
\end{equation}
which agrees with Eq.~(\ref{eq:kompaneets}) without the bulk velocity effect $\delta_v$. 

\subsection{Including bulk motion}
\label{sec:correction_bulk}
In the above derivation, we have used the Maxwell distribution for nucleons. Next, let us include the bulk motion. 
The distribution function of nucleons including bulk motion is given by $g_N({\bm p})=n_N(2\pi m_N k_BT)^{-3/2}e^{-({\bm p}-m_N {\bm V})^2/2m_Nk_BT}$.
Then, the bulk velocity of a fluid, ${\bm V}$, is given by
\begin{equation}
{\bm V}=\left<\frac{{\bm p}}{m_N}\right>\equiv \frac{\int d^3p \left(\displaystyle\frac{{\bm p}}{m_N}\right)g_N({\bm p})}{\int d^3p~ g_N({\bm p})}\,,
\end{equation}
where $\langle\dots \rangle$ denotes the average over the momentum distribution of  nucleons.
The second moment of ${\bm p}/m_N$ is
\begin{equation}
\left<\frac{p^2}{m_N^2}\right>=\bracket{v^2}+V^2\,,
\end{equation}
where ${\bm v}\equiv{\bm p}/m_N-{\bm V}$ is the velocity of a nucleon
in the rest frame of a fluid element. Assuming that a nucleon's momentum
distribution in the rest frame of a fluid element is a Maxwell
distribution, we have $\bracket{v^2}=3k_BT/m_N$.
Including the bulk velocity in Eq. (\ref{eq:kompaneets-wo-v}), we obtain
\begin{equation}
\frac{d f}{d y}=
\frac{1}{x^2}\frac{\partial}{\partial x}
\left\{
  x^6
  \left[
    f-f^2+(1+\delta_v)f'
  \right]
\right\}\,,
\label{eq:kompaneets_v2}
\end{equation}
which agrees with Eq.~(\ref{eq:kompaneets}).

\end{document}